\documentstyle[11pt]{article}
\begin{document}

\begin{flushright}
{hep-ph/0101081}\\

\end{flushright}
\vspace{1cm}
\begin{center}
{\Large \bf  $|\Delta B|=1$ Weak Effective Lagrangian \\
\vspace{0.5cm}
in the Minimal Flavor Violation Supersymmetry}
\vspace{.2cm}
\end{center}
\vspace{1cm}
\begin{center}
{Tai-Fu Feng$^{a,b,d}$\hspace{0.5cm}Xue-Qian Li$^{a,b,d}$}
\hspace{.5cm}Guo-Li Wang$^{a,c,d}$\\
\vspace{.5cm}

{$^a$CCAST (World Laboratory), P.O.Box 8730,
Beijing 100080, P. R. China}\\
{$^b$Department of Physics, Nankai
University, Tianjin 300070, P. R. China}\\
{$^c$Department of Physics, Fujian Normal
University, Fuzhou, 350007, P. R. China}\\
{$^d$Institute of Theoretical Physics, Academia Sinica, P.O. Box
2735, Beijing 100080, P. R. China}\\
\vspace{.5cm}

\end{center}
\hspace{3in}

\begin{center}
\begin{minipage}{11cm}
{\large\bf Abstract}

{\small To evaluate the weak decays of b-hadrons, the
$\Delta B=1$ weak effective Lagrangian is the foundation.
Any new physics beyond the
standard model (SM) would contribute to the effective
Lagrangian through the loop integration at the weak
scale and evolution from the weak scale down to the
hadronic scale. In this work we present a systematic
analysis on the effective Lagrangian
which mediates hadronic
$|\Delta B|=1$ processes in the framework of the minimal
flavor violation supersymmetry as well as a numerical
evaluation of the Wilson coefficients in the effective
theory.}
\end{minipage}
\end{center}

\vspace{4mm}
\begin{center}
\end{center}


\section{Introduction}

The forthcoming B-factories will make more precise measurements on the
rare B-decay processes and those measurements would set more strict
constraints on the new physics beyond the SM.
The purpose to investigate B-decays, especially the rare decay
modes is to search for traces of new physics and determine its
parameter space.

The new physics effects on the rare B processes
are intensively discussed in literature.
If we believe that the SM is only an effective theory
and the supersymmetry is  more fundamental,
measurements of rare
B-processes will definitely enrich our knowledge in
this field. But before we can really pin down any new physics
effects, we need to carry out a thorough exploration in this area,
not only in SM, but also in many plausible models, especially the
supersymmetric model.

The calculation of the rate of inclusive
decay $B\rightarrow X_s\gamma$ is presented by authors of \cite{ciuchini1,
ciafaloni, borzumati} in the two-Higgs doublet model (2HDM).
The supersymmetric effect on $B\rightarrow X_s\gamma$ is discussed
in \cite{bertolini, barbieri, borzumati1} and the
next-to-leading order (NLO)
QCD corrections are given in \cite{ciuchini2}. The transition
$b\rightarrow s\gamma\gamma$ in the supersymmetric extension
of the standard model is computed in \cite{bertolini1}. The hadronic
B decays\cite{cottingham} and CP-violation in those processes\cite{barenboim}
have been discussed also. The authors of \cite{hewett} have discussed
possibility of observing supersymmetric effects in
rare decays $B\rightarrow X_s\gamma$ and $B\rightarrow X_se^+e^-$
at the B-factory. Studies on decays $B\rightarrow
(K, K^*)l^+l^-$ in SM and supersymmetric model have been carried out
in \cite{ali}. The supersymmetric effects on these processes are very
interesting and studies on them may shed some light on the
general characteristics of the supersymmetric model.
A relevant review can be found in \cite{masiero}.
For oscillations of $B_{0}-\overline{B}_{0}$ ($K_{0}-\overline{K}_{0}$),
calculations have been done in the
SM and 2HDM. As for the supersymmetric
extension of SM, the calculation
involving the gluino contributions
should be re-studied carefully for gluino has a nonzero mass.
At the NLO approximation, the QCD corrections to the
$B_0-\overline B_0$ mixing in the supersymmetry model have been discussed
recently. The authors of \cite{ciuchini3, contino} applied the
mass-insertion method to estimate QCD corrections to
the $B_0-\overline B_0$ mixing.
The calculations including the gluon-mediated QCD were given in\cite{Soff},
and later we have re-derived the formulation by including the contribution of gluinos
\cite{Feng1}.
%%%%%%%%%%%%%%%%%%%%%%%%%%%%%%%%%%%%%%%%%%%%%%%%%%%%%%%%%%
%For the two body decay modes $b\rightarrow
%sg$ and $b\rightarrow s\gamma$, a complete
%next-to-leading logarithm (NLO) prediction is performed
%recently in the SM case\cite{Greub,Greub1}.
%%%%%%%%%%%%%%%%%%%%%%%%%%%%%%%%%%%%%%%%%%%%%%%%%%%%%%%%%%

The supersymmetry effects influence the rare B processes
in two ways:
\begin{itemize}
\item the Wilson coefficients of the operators existing in the
standard model case receive corrections from the supersymmetry
sector. As far as the four-quark operators are concerned, the
supersymmetric contribution begins at the order of $O(\alpha_s)$ along with
the SM QCD corrections.
\item when the supersymmetry effects are taken into account,
the operator basis is enlarged, new operators emerge.
\end{itemize}
Generally speaking, theoretical predictions on the inclusive decay rates of
B-mesons rest on solid grounds due to the fact that these rates can be
systematically expanded in powers of $\frac{\Lambda_{QCD}}{m_b}$\cite{
Bigi1,Bigi2}, where the leading term corresponds to the decay width of
free b-quark. As the power corrections only start at
${\cal O}(\frac{\Lambda_{QCD}^2}{m_b^2})$, they affect these rates by at
most a few percent. Theoretically, the non-spectator effects of order $16\pi^2
\Big(\frac{\Lambda_{QCD}}{m_b}\Big)^3$  could be large\cite{Bigi3,Neubert},
especially for the charmless decay modes of B-mesons
\cite{Neubert,dai,Kroll}.
The main contributions to the lifetimes of B-mesons and
$\Lambda_b$ are from the b-quark decays which are thoroughly
studied in the framework of SM.

All the theoretical calculations are based on the weak effective
Lagrangian which determines the effective vertices in the concerned
Feynman diagrams\cite{Altarelli}.
The NLO calculation
has been carried out in SM\cite{Lenz,Lenz1}, but is not
complete for the supersymmetric extension.
In order to study the supersymmetry effects in those low energy
processes, one should obtain a complete effective Lagrangian which
includes the supersymmetric contributions at the order
$O(\alpha_s)$. Here we consider
the supersymmetric model with minimal flavor violation (MFV),
i.e. all flavor transitions occur only in the charged-current sector and
are determined by the standard
Cabibbo-Kobayashi-Maskawa (CKM) mechanism.

In this work, we perform a complete analysis on the $|\Delta B|=1$
effective Lagrangian, including the current-current operators and
penguin-induced operators within the framework of the
MFV supersymmetry.

After matching between the full MFV
supersymmetric theory and the effective Lagrangian,
the Wilson coefficients for concerned operators are
obtained at the weak scale. Using the recently developed
two-loop QCD anomalous dimension matrix of flavor
changing four-quark operators\cite{Buras1}, we discuss the evolution of
the $|\Delta B|=1$ non-leptonic effective Lagrangian from the weak scale
down to the hadronic scale.

The paper is organized as follows. In section 2, we review
the minimal flavor violation supersymmetric model and give
the notations adopted in our analysis. In section 3, the detailed
derivations of the effective Lagrangian at the weak and hadronic
scales are made. Then in section 4, we present the numerical results
which explicitly demonstrate the difference of the Wilson
coefficients in MFV supersymmetry and SM. We discuss
the numerical results and make a short summary
in section 5. Some complicated and tedious
expressions are collected in the appendices.

\section{The supersymmetry with minimal flavor violation \label{notation}}

Throughout this paper we adopt the notation similar to Ref.\cite{rosiek},
the expressions of the concerned propagators and vertices can be found
in the appendix
of Ref.\cite{rosiek}. For convenience, we write down the superpotential
and relevant mixing matrices. The most general form of the
superpotential which does not violate gauge invariance and the
SM conservation laws is
\begin{eqnarray}
&&{\cal W}=\mu \epsilon_{ij}\hat{H}_{i}^{1}\hat{H}_{j}^{2}+
\epsilon_{ij}h_{l}^{I}\hat{H}_{i}^{1}\hat{L}^{I}_{j}\hat{R}^{I}-
h_{d}^{I}(\hat{H}_{1}^{1}\hat{Q}^{I}_{2}-\hat{H}_2^1V^{IJ}
\hat{Q}_1^J)\hat{D}^{I}-
h_{u}^{I}(\hat{H}_{1}^{2}V^{*JI}\hat{Q}^{J}_{2}-
\hat{H}_2^2\hat{Q}_1^I)\hat{U}^{I}.
\label{superpotential}
\end{eqnarray}
Here, the weak ${\rm SU}(2)$ doublets of quark superfields have
been written in the form
$$\left(\begin{array}{c}
V^{IJ}\hat{Q}_1^J \\ \hat{Q}_2^I \end{array}\right)\;,$$
where $I=1,\;2,\;3$ are the indices of generations.
The Higgs and lepton doublets are denoted by  $\hat{H}^{1}$,
$\hat{H}^{2}$ and $\hat{L}^{I}$, respectively.
The rest superfields $\hat{U}^{I}$, $\hat{D}^{I}$ and $\hat{R}^{I}$
are quark superfields of u- and d-types and  charged leptons
in singlets of the weak SU(2). Indices i, j are contracted
for the SU(2) group, and $h_{l}$, $h_{u,d}$
are the Yukawa couplings.

In order to break the supersymmetry,
the soft breaking terms are introduced as
\begin{eqnarray}
&&{\cal L}_{soft}=-m_{H^1}^2H_i^{1*}H_i^1-m_{H^2}^2H_i^{2*}H_i^2
-m_{L^I}^2\tilde{L}_i^{I*}\tilde{L}_i^{I}\nonumber \\
&&\hspace{2.cm}-m_{R^I}^2\tilde{R}^{I*}\tilde{R}^{I}
-m_{Q^I}^2\tilde{Q}_i^{I*}\tilde{Q}_i^{I}
-m_{U^I}^2\tilde{U}^{I*}\tilde{U}^{I}\nonumber \\
&&\hspace{2.cm}-m_{D^I}^2\tilde{D}^{I*}\tilde{D}^{I}
+(m_1\lambda_B\lambda_1+m_2\lambda_A^i\lambda_A^i\nonumber \\
&&\hspace{2.cm}+m_3\lambda_G^a\lambda_G^a+h.c.)+\Big[B\mu\epsilon_{ij}H_i^1H_j^2
+\epsilon_{ij}A_l^Ih_{l}^{I}H_{i}^{1}\tilde{L}^{I}_{j}\tilde{R}^{I}
\nonumber \\
&&\hspace{2.cm}-A_d^Ih_{d}^{I}(H_{1}^{1}\tilde{Q}^{I}_{2}
-H_2^1V^{IJ}\tilde{Q}_1^J)\tilde{D}^{I}-
A_u^Ih_{u}^{I}(H_{1}^{2}V^{*JI}\tilde{Q}^{J}_{2}-
H_2^2\tilde{Q}_1^I)\tilde{U}^{I}+h.c.\Big]+\cdots
\label{soft}
\end{eqnarray}
where $m_{H^1}^2,\;m_{H^2}^2,\;m_{L^I}^2,\;m_{R^I}^2,\;m_{Q^I}^2,\;
m_{U^I}^2$ and $m_{D^I}^2$ are the parameters of dimension two,
while $m_3,\;m_2,\;m_1$ denote the masses of $\lambda_G^a\;(a=1,\;2,\;
\cdots\;8),\;\lambda_A^i\;(i=1,\;2,\;3)$, and $\lambda_B$, the $SU(3)\times
SU(2)\times U(1)$ gauginos. $B$ is a free parameter of dimension one.
$A_{l}^{I},\;A_{u}^{I},\;A_{d}^{I}\;(I=1,\;2,\;3)$ are the soft breaking
trilinear couplings of scalars. The dots in Eq.\ref{soft}
stand for flavor-off-diagonal terms (e.g. $m_{Q^K}^2Q_1^{I*}
Q_1^J(\delta^{KI}\delta^{KJ}-V^{KI*}V^{KJ})$) that are assumed to
be negligible. Such terms do occur in our numerical calculation
in section 4, but are indeed very small.
%parameters that result in mass splitting between the leptons, quarks and
%their supersymmetric partners.

Taking into account of the soft breaking
terms Eq.(\ref{soft}), we can study the
phenomenology within the supersymmetric extension of
the standard model with minimal flavor violation
(MFV MSSM). Once the favor-off-diagonal soft supersymmetry
breaking terms are neglected, the squark mass matrices
can be written as $2\times 2$ matrices for each flavor
separately:
\begin{eqnarray}
&&m_{\tilde{U}^I}^2=\left(
\begin{array}{cc}
m_{Q^I}^2+m_{u^I}^2+(\frac{1}{2}-\frac{2}{3}\sin^2\theta_{\rm W})
\cos 2\beta m_{\rm Z}^2 & -m_{u^I}(A_u^I+\mu\cot\beta) \\
-m_{u^I}(A_u^I+\mu\cot\beta) & m_{U^I}^2+m_{u^I}^2+
\frac{2}{3}\sin^2\theta_{\rm W}\cos 2\beta m_{\rm Z}^2
\end{array} \right),
\label{stmass}
\end{eqnarray}
and
\begin{eqnarray}
&&m_{\tilde{D}^I}^2=\left(
\begin{array}{cc}
m_{Q^I}^2+m_{d^I}^2+(\frac{1}{2}+\frac{1}{3}\sin^2\theta_{\rm W})
\cos 2\beta m_{\rm Z}^2 & -m_{d^I}(A_d^I+\mu\tan\beta) \\
-m_{d^I}(A_d^I+\mu\tan\beta) & m_{D^I}^2+m_{d^I}^2-
\frac{1}{3}\sin^2\theta_{\rm W}\cos 2\beta m_{\rm Z}^2
\end{array} \right),
\label{sbmass}
\end{eqnarray}
with $m_{u^I},\;m_{d^I}\;(I=1,\;2,\;3)$ are the masses of
the I-th generation quarks.

The SM and the MSSM differ in their Higgs sectors.
There are four charged scalars,
two of them are physical massive Higgs bosons and other are
massless Goldstones. The mixing matrix can be written as:
\begin{equation}
{\cal Z}_{H}=\left(
\begin{array}{cc}
\sin\beta & -\cos\beta \\
\cos\beta & \sin\beta \end{array}
\right) \label{zh} \end{equation}
with $\tan\beta=\frac{\upsilon_{2}}{\upsilon_{1}}$ and $v_1,v_2$ are
the vacuum expectation values of the two Higgs scalars.
Another matrix that we will use is the chargino mixing
matrix. The supersymmetric partners of the charged Higgs and $W^{\pm}$
combine to give two Dirac fermions: $\chi^{\pm}_{1}$,
$\chi^{\pm}_{2}$. The two mixing matrices ${\cal Z}^{\pm}$
appearing in the Lagrangian are defined as
\begin{eqnarray}
&&({\cal Z}^{-})^{T}{\cal M}_{c}{\cal Z}^{+} = diag(m_{\chi_{1}},
m_{\chi_{2}}),
\label{zpm}
\end{eqnarray}
where
$${\cal M}_{c}=\left(\begin{array}{cc}
2m_2&{1\over \sqrt{2}}g_2\upsilon_2\\
{1\over \sqrt{2}}g_2\upsilon_1&\mu
\end{array}\right)$$
is the mass matrix of charginos with $g_{2}$ denoting
the gauge coupling of ${\rm SU}(2)$.
In a similar way, ${\cal  Z}_{U,D}$ diagonalize the mass
matrices of the up- and down-type squarks respectively:
\begin{eqnarray}
&& {\cal Z}_{{\tiny U}^I}^{\dag}m^2_{\tilde{U}^I}
{\cal Z}_{{\tiny U}^I}=
diag(m^2_{\tilde{{\tiny U}}^I_1},m^2_{\tilde{{\tiny U}}^I_2})\;,
\nonumber \\
&& {\cal Z}_{{\tiny D}^I}^{\dag}m^2_{\tilde{D}^I}
{\cal Z}_{{\tiny D}^I}=
diag(m^2_{\tilde{{\tiny D}}^I_1},m^2_{\tilde{{\tiny D}}^I_2})\;.
\label{zud}
\end{eqnarray}
With those mixing matrices defined above, we can write the interaction
vertices as in Ref.\cite{rosiek}.

\section{Matching  the coefficients of operators}

As in the SM case, we need to obtain the low-energy effective
Lagrangian with five quarks, and while deriving it,
the heavy supersymmetric
degrees of freedom as well as that of SM, including  top
quark, W-bosons, charged Higgs
bosons and the supersymmetric partners of the standard particles
are integrated out. In this work we
only retain the operators up to dimension six. In this
approximation the effective Lagrangian for $|\Delta B|=1$ reads
\begin{eqnarray}
&&{\cal L}_{eff}=-\frac{4G_F}{\sqrt{2}}V_{ts}^*V_{tb}
\Big[\sum\limits_{i=1}^{10}\Big(C_i^c(\mu)Q_i^c
+\tilde{C}_i^c(\mu)\tilde{Q}
_i^c\Big)+\sum\limits_{j=1}^5C_j^p(\mu)Q_j^p\Big]
\label{hamilton}
\end{eqnarray}
where $G_F$ is the Fermi coupling constant, $C_i^{c}(\mu),\;\tilde{C}_i^c
(\mu),\;C_j^{p}(\mu)\;(i=1,\;2,\;\cdots,\;10;\;\;j=1,\;\cdots,\;5)$
are the Wilson coefficients evaluated at the scale $\mu$; $V_{tb}$ and
$V_{ts}$ are the matrix elements of the CKM
matrix. Making the effective Lagrangian close
under the QCD renormalization, we include the penguin operator
$Q_5^p$ beside those four-quark operators.

As commonly adopted in literature, we classify the operators as
the current-current operators which are originally
induced by the tree level W-exchange interaction
and one-loop 'box' diagrams, and the
"penguin-"induced operators.
Both of them would undergo QCD corrections and receive
contributions from supersymmetric particles via loops.

The current-current operators are written as\cite{Buras1}
\begin{eqnarray}
&&Q_1^c=\Big(\overline{s}_\alpha\gamma_\mu\omega_-c_\beta\Big)
\Big(\overline{c}_\beta\gamma^\mu\omega_-b_\alpha\Big)\;,\nonumber \\
&&Q_2^c=\Big(\overline{s}_\alpha\gamma_\mu\omega_-c_\alpha\Big)
\Big(\overline{c}_\beta\gamma^\mu\omega_-b_\beta\Big)\;,\nonumber \\
&&Q_3^c=\Big(\overline{s}_\alpha\omega_-c_\beta\Big)
\Big(\overline{c}_\beta\omega_-b_\alpha\Big)\;,\nonumber \\
&&Q_4^c=\Big(\overline{s}_\alpha\omega_-c_\alpha\Big)
\Big(\overline{c}_\beta\omega_-b_\beta\Big)\;,\nonumber \\
&&Q_5^c=\Big(\overline{s}_\alpha\sigma_{\mu\nu}\omega_-c_\beta\Big)
\Big(\overline{c}_\beta\sigma^{\mu\nu}\omega_-b_\alpha\Big)\;,\nonumber \\
&&Q_6^c=\Big(\overline{s}_\alpha\sigma_{\mu\nu}\omega_-c_\alpha\Big)
\Big(\overline{c}_\beta\sigma^{\mu\nu}\omega_-b_\beta\Big)\;,\nonumber \\
&&Q_7^c=\Big(\overline{s}_\alpha\gamma_\mu\omega_-c_\beta\Big)
\Big(\overline{c}_\beta\gamma^\mu\omega_+b_\alpha\Big)\;,\nonumber \\
&&Q_8^c=\Big(\overline{s}_\alpha\gamma_\mu\omega_-c_\alpha\Big)
\Big(\overline{c}_\beta\gamma^\mu\omega_+b_\beta\Big)\;,\nonumber \\
&&Q_9^c=\Big(\overline{s}_\alpha\omega_-c_\beta\Big)
\Big(\overline{c}_\beta\omega_+b_\alpha\Big)\;,\nonumber \\
&&Q_{10}^c=\Big(\overline{s}_\alpha\omega_-c_\alpha\Big)
\Big(\overline{c}_\beta\omega_+b_\beta\Big)\;,
\label{current}
\end{eqnarray}
where $\sigma_{\mu\nu}=\frac{i}{2}[\gamma_\mu,\gamma_\nu]$ and
$\omega_\pm=\frac{1\pm\gamma_5}{2}$. In the
standard model, there are only two such operators i.e. $Q_1^c,\;
Q_2^c$, 18 new operators
are induced when supersymmetry  takes part in the game.
There are other ten current-current operators which are simply
obtained by interchanging  $\omega_\pm\leftrightarrow\omega_{\mp}$
in $Q_i^c$, i.e.
$\tilde{Q}_i^c=Q_i^c(\omega_\pm\leftrightarrow\omega_\mp)$.
Due to the small CKM entries for the u-analog operators in the
$|\Delta B|=1$ effective Lagrangian,
$V_{us}^*V_{ub}<<V_{cs}^*V_{cb}$ the
u-quark analogs of the effective Lagrangian in Eq.\ref{hamilton}
can be neglected.
The basis of the penguin-induced operators consists of
\cite{Gilman,Buras3,Buras4,Ciuchini,
Chetyrkin}
\begin{eqnarray}
&&Q_1^p=\Big(\overline{s}_\alpha\gamma_\mu\omega_-b_\alpha\Big)
\sum\limits_q\Big(\overline{q}_\beta\gamma^\mu\omega_-q_\beta\Big)\;,
\nonumber \\
&&Q_2^p=\Big(\overline{s}_\alpha\gamma_\mu\omega_-b_\beta\Big)
\sum\limits_q\Big(\overline{q}_\beta\gamma^\mu\omega_-q_\alpha\Big)\;,
\nonumber \\
&&Q_3^p=\Big(\overline{s}_\alpha\gamma_\mu\omega_-b_\alpha\Big)
\sum\limits_q\Big(\overline{q}_\beta\gamma^\mu\omega_+q_\beta\Big)\;,
\nonumber \\
&&Q_4^p=\Big(\overline{s}_\alpha\gamma_\mu\omega_-b_\beta\Big)
\sum\limits_q\Big(\overline{q}_\beta\gamma^\mu\omega_+q_\alpha\Big)\;,
\nonumber \\
&&Q_5^p=\frac{1}{(4\pi)^2}\bar{s}g_sG\cdot \sigma(m_s\omega_-
+m_b\omega_+)b\;,
\label{penguin}
\end{eqnarray}
with $q=u,\;d,\;c,\;s,\;b$, $G_{\mu\nu}\equiv G_{\mu\nu}^aT^a$
denotes the gluon field strength
tensor, $G_{\mu\nu}^a=\partial_{\mu}G_{\nu}^a-\partial_{\nu}
G_{\mu}^a+g_sf^{abc}G_{\mu}^bG_{\nu}^c$, and $G\cdot \sigma\equiv
G_{\mu\nu}\sigma^{\mu\nu}$.

In the following sections,
we will derive the Wilson coefficients for the current-current
operators in Eq.\ref{current} and penguin-induced operators in Eq.\ref{penguin}.

%%%%%%%%%%%%%%%%%%%%%%%%%%%%%%%%%%%%%%%%%%%%%%%%%%%%%%%%%%%%%
\subsection{The difference of quark field normalization
in the full and effective theories
while taking the $\overline{\rm MS}$ scheme }

In order to systematically investigate
all corrections of the supersymmetry QCD to the vertex $\bar duW$ including
the self-energies of external quark legs, Ciuchini
{\it et.al} used the on-shell scheme to subtract the
divergence of the quark fields and the $\bar du{\rm W}$ vertex
that originates from supersymmetric partners,
whereas the divergence originating from the SM sector is still
subtracted out in the $\overline{\rm MS}$-scheme\cite{ciuchini2}.
In this work, we will employ the $\overline{\rm MS}$ scheme
throughout and show that the results are qualitatively
consistent with theirs.

It is noted that there is a difference of the normalization of the quark fields
in the full supersymmetric theory and the effective theory
while taking the $\overline{\rm MS}$ scheme.
As a matter of fact, for the self-energy which determines the
renormalization of the wave functions, the Feynman diagrams in the
effective theory are the same as the standard model part of the
full theory, therefore there is not normalization ambiguity in the
SM case.
When the supersymmetry sector is included, there exists an extra
term from the supersymmetry sector and the normalizations of the external
quark fields are not the same after renormalization in the $\overline{\rm MS}$
scheme, as
$${1\over 1+\Delta Z_{SM}^{full}+\Delta Z_{MSSM}^{full}}\neq
{1\over 1+\Delta Z^{eff}}\;\;,$$
where $\Delta Z_{SM}^{full}=\Delta Z^{eff}$ and the superscript "eff"
denotes the quantities in the effective theory.
Thus one cannot simply match the vertex-induced Lagrangian in the
full and effective theories because the external quark fields have
different normalizations. That is understood  that in both
the full and effective theories, the SM quarks exist, but
the supersymmetry particles (squarks and gluinos) only exist in the full
theory, but are integrated out to produce the effective  Lagrangian.
The difference
 $\Delta Z^{full}-\Delta Z^{eff}=
\Delta Z_{MSSM}^{full}$ is a finite renormalization effect which
should be included when we match the Lagrangian in the full and
effective theories. Namely, when we match the Lagrangians, we not
only consider the contributions from the vertices, but also
include this normalization difference. We find that
the large logarithms which exist
in the vertex contributions and this normalization
difference would cancel each other exactly and then the decoupling
theorem is obvious.

The difference manifests as a finite renormalization
contribution to the $\Delta B=1$ effective
Lagrangian, which will be expressed explicitly in the following computation.

The supersymmetric contributions to the self-energy are
\begin{eqnarray}
&&i\Sigma^{^{susy}}_u(p)=-i\frac{\alpha_s}{4\pi}C_F\Bigg\{\bigg(
\Delta+\ln x_\mu+\frac{3}{2}+\frac{x_{_{\tilde{U}_i^I}}}{x_{_{\tilde{g}}}
-x_{_{\tilde{U}_i^I}}}+\frac{x_{_{\tilde{U}_i^I}}\ln
x_{_{\tilde{U}_i^I}}}{x_{_{\tilde{g}}}-x_{_{\tilde{U}_i^I}}}
-\frac{x_{_{\tilde{g}}}^2\ln x_{_{\tilde{g}}}-x_{_{\tilde{g}}}
x_{_{\tilde{U}_i^I}}\ln x_{_{\tilde{U}_i^I}}}
{(x_{_{\tilde{g}}}-x_{_{\tilde{U}_i^I}})^2}\bigg)
\Big(Z_{\tilde{U}^I}^{1i}Z_{\tilde{U}^I}^{1i*}/\!\!\!p\omega_-
\nonumber \\
&&\hspace{2.5cm}
+Z_{\tilde{U}^I}^{2i}Z_{\tilde{U}^I}^{2i*}/\!\!\!p\omega_+\Big)
+2m_{\tilde{g}}\frac{x_{_{\tilde{U}_i^I}}(\ln x_{_{\tilde{g}}}-
\ln x_{_{\tilde{U}_i^I}})}{x_{_{\tilde{g}}}-x_{_{\tilde{U}_i^I}}}
\Big(Z_{\tilde{U}^I}^{1i*}Z_{\tilde{U}^I}^{2i}\omega_-
+Z_{\tilde{U}^I}^{1i}Z_{\tilde{U}^I}^{2i*}\omega_+\Big)
\Bigg\}\;,\nonumber \\
&&i\Sigma^{^{susy}}_d(p)=-i\frac{\alpha_s}{4\pi}C_F\Bigg\{\bigg(
\Delta+\ln x_\mu+\frac{3}{2}+\frac{x_{_{\tilde{D}_i^I}}}{x_{_{\tilde{g}}}
-x_{_{\tilde{D}_i^I}}}+\frac{x_{_{\tilde{D}_i^I}}\ln
x_{_{\tilde{D}_i^I}}}{x_{_{\tilde{g}}}-x_{_{\tilde{D}_i^I}}}
-\frac{x_{_{\tilde{g}}}^2\ln x_{_{\tilde{g}}}-x_{_{\tilde{g}}}
x_{_{\tilde{D}_i^I}}\ln x_{_{\tilde{D}_i^I}}}
{(x_{_{\tilde{g}}}-x_{_{\tilde{D}_i^I}})^2}\bigg)
\Big(Z_{\tilde{D}^I}^{1i}Z_{\tilde{D}^I}^{1i*}/\!\!\!p\omega_-
\nonumber \\
&&\hspace{2.5cm}
+Z_{\tilde{D}^I}^{2i}Z_{\tilde{D}^I}^{2i*}/\!\!\!p\omega_+\Big)
+2m_{\tilde{g}}\frac{x_{_{\tilde{D}_i^I}}(\ln x_{_{\tilde{g}}}-
\ln x_{_{\tilde{D}_i^I}})}{x_{_{\tilde{g}}}-x_{_{\tilde{D}_i^I}}}
\Big(Z_{\tilde{D}^I}^{1i*}Z_{\tilde{D}^I}^{2i}\omega_-
+Z_{\tilde{D}^I}^{1i}Z_{\tilde{D}^I}^{2i*}\omega_+\Big)
\Bigg\}\;,
\label{susyse}
\end{eqnarray}
where $x_{\mu}={\mu_W^2\over m_{\rm W}^2}\;,\;
x_i=\frac{m_i^2}{m_{\rm W}^2}\;.$
In the expressions the first term
is the correction to the wave function
of the quarks whereas the second term corresponds to the
supersymmetric contributions to the quark masses.
In Eq.\ref{susyse}, if we complete the renormalization for the
quark fields in the $\overline{\rm MS}$-scheme, the divergent part
in the renormalization multiplier which should be subtracted is
\begin{equation}
\delta Z_q^{\pm}(\overline{\rm MS})=\frac{\alpha_s}{4\pi}C_F\Delta \;.
\label{rems}
\end{equation}
When matching the full and effective theories in
$\overline{\rm MS}$ scheme, there is a difference of the renormalized quark
fields, so that we need to derive the extra contribution.

In literature \cite{ciuchini2}, the on-mass-shell
renormalization \cite{Denner} is adopted, by which
after renormalization the residue of the
self-energy should be one at the physical mass, so that the
normalization ambiguity does not exist.  However, in
the $\overline{\rm MS}$ scheme, there is no such a requirement.
From Eq.\ref{susyse}, we have the difference of the normalization
of the quark fields in the full supersymmetric
theory and the effective theory while taking the $\overline{\rm MS}$ scheme as
\begin{eqnarray}
&&\Delta Z_{u^I}^{-}=\frac{\alpha_s}{4\pi}C_F\bigg(
\ln x_\mu+\frac{3}{2}+\sum\limits_{i}
Z_{\tilde{U}^I}^{1i}Z_{\tilde{U}^I}^{1i*}\Big(
\frac{x_{_{\tilde{U}_i^I}}}{x_{_{\tilde{g}}}
-x_{_{\tilde{U}_i^I}}}+\frac{x_{_{\tilde{U}_i^I}}\ln
x_{_{\tilde{U}_i^I}}}{x_{_{\tilde{g}}}-x_{_{\tilde{U}_i^I}}}
%\nonumber \\
%&&\hspace{3.1cm}
-\frac{x_{_{\tilde{g}}}^2\ln x_{_{\tilde{g}}}-x_{_{\tilde{g}}}
x_{_{\tilde{U}_i^I}}\ln x_{_{\tilde{U}_i^I}}}
{(x_{_{\tilde{g}}}-x_{_{\tilde{U}_i^I}})^2}\Big)\bigg)
\nonumber\; ,\\
&&\Delta Z_{d^I}^{-}=\frac{\alpha_s}{4\pi}C_F\bigg(
\ln x_\mu+\frac{3}{2}+\sum\limits_{i}
Z_{\tilde{D}^I}^{1i}Z_{\tilde{D}^I}^{1i*}\Big(
\frac{x_{_{\tilde{D}_i^I}}}{x_{_{\tilde{g}}}
-x_{_{\tilde{D}_i^I}}}+\frac{x_{_{\tilde{D}_i^I}}\ln
x_{_{\tilde{D}_i^I}}}{x_{_{\tilde{g}}}-x_{_{\tilde{D}_i^I}}}
-\frac{x_{_{\tilde{g}}}^2\ln x_{_{\tilde{g}}}-x_{_{\tilde{g}}}
x_{_{\tilde{D}_i^I}}\ln x_{_{\tilde{D}_i^I}}}
{(x_{_{\tilde{g}}}-x_{_{\tilde{D}_i^I}})^2}\Big)\bigg)\;,
\label{difnorm}
\end{eqnarray}
When we match the Lagrangians in the full and effective theories,
we not only consider the vertex-induced contributions, but also
need to include this finite renormalization contribution. Thus
taking this normalization difference into account,
we obtain the Wilson coefficients of the current-current
operators in the full and effective theories.

%%%%%%%%%%%%%%%%%%%%%%%%%%%%%%%%%%%%%%%%%%%%%%%%%%%%%%%%%%%%%%%%%
\subsection{The Wilson coefficients of current-current operators}

\begin{figure}
\setlength{\unitlength}{1mm}
\begin{center}
\begin{picture}(230,200)(55,90)
\put(50,30){\includegraphics{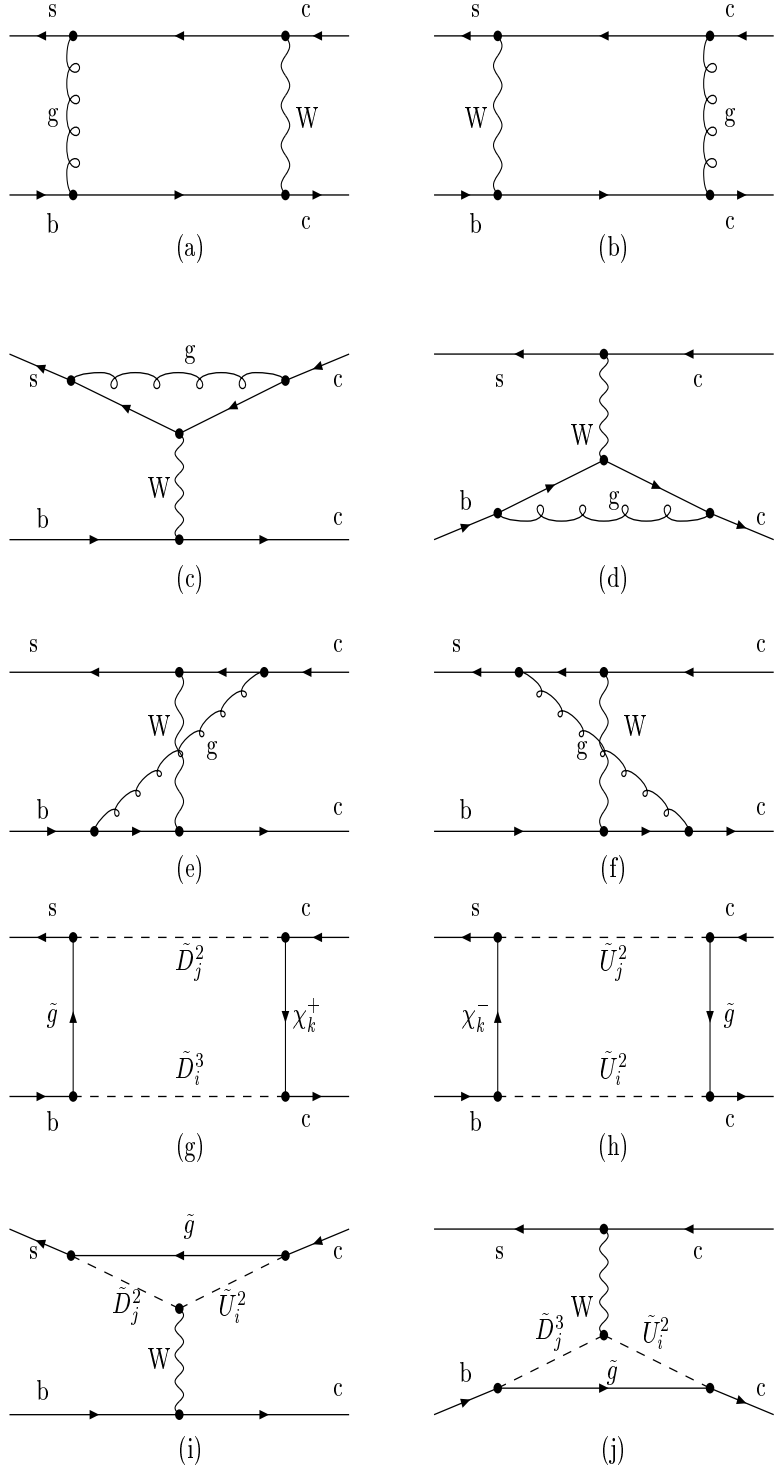}}
\end{picture}
\caption[]{The one-loop Feynman diagrams
in the minimal flavor violation supersymmetry for the
current-current
operators in the full theory at the weak energy scale}
\label{fig1}
\end{center}
\end{figure}

\begin{figure}
\setlength{\unitlength}{1mm}
\begin{center}
\begin{picture}(230,200)(55,90)
\put(50,-50){\includegraphics{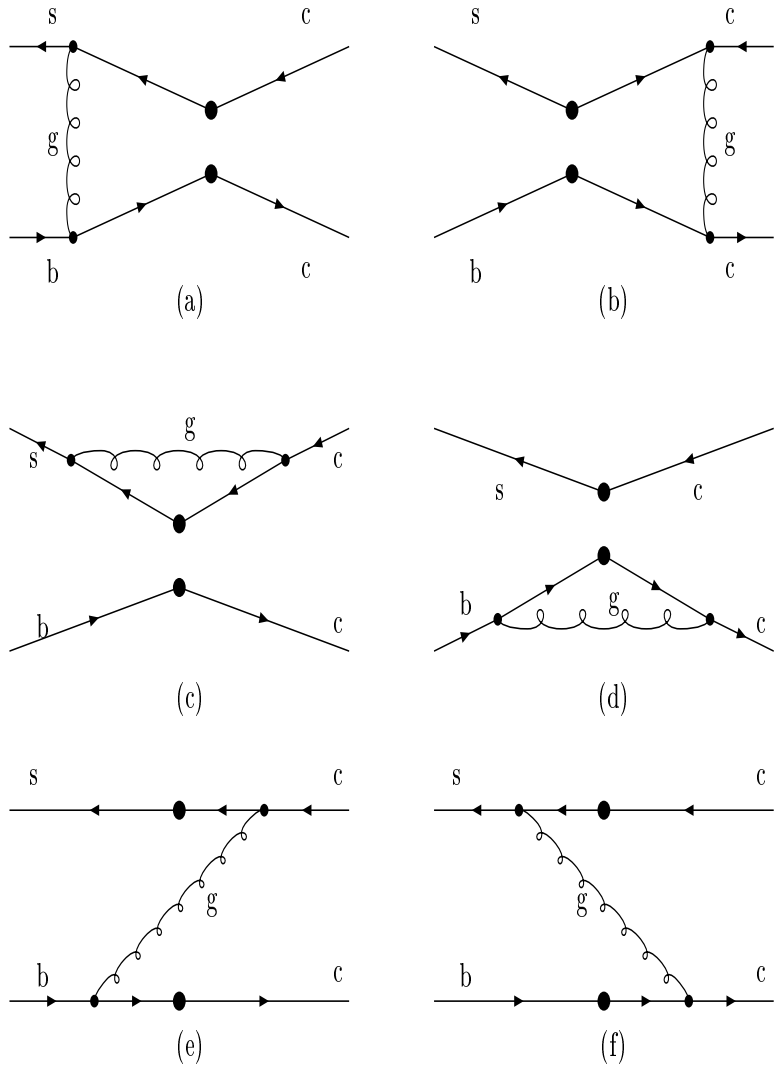}}
\end{picture}
\caption[]{The Feynman diagrams for QCD-corrections to the
current-current operators in effective theory with five
quarks}
\label{fig2}
\end{center}
\end{figure}

The Wilson coefficients $C_i^c,\;\tilde{C}_i^c\;(i=1,\;2,\;\cdots,\;10)$
in Eq.\ref{hamilton} can be determined by the requirement that the amplitude
$A_{full}$ in the full theory is equal to the corresponding amplitude in the
effective theory at the weak scale
\begin{eqnarray}
&&A_{full}=A_{eff}=\frac{4G_F}{\sqrt{2}}V_{cs}^*V_{cb}\sum_{i=1}^{10}\Big\{
C_i^cQ_i^{c}+\tilde{C}_i^c\tilde{Q}_i^{c}-\Big[\Delta Z_{c}^{-}+
\frac{1}{2}\Delta Z_{b}^-+\frac{1}{2}\Delta Z_{s}^-
\Big]Q_2^c\Big\}.
\label{match}
\end{eqnarray}
The QCD induced one-loop Feynman diagrams responsible for $\Delta B=1$
effective Lagrangian in the minimal flavor violation supersymmetric
theory and effective theory are drawn in Fig.\ref{fig1}
and Fig.\ref{fig2} respectively.
The last term of Eq.\ref{match} originates from the difference of
quark field normalization in the full and effective
theories. Considering the QCD corrections of
current-current operators, we can extract the Wilson
coefficients $C_i^c,\;\tilde{C}_i^c$ at the $\mu_{\rm W}$
scale. The expressions for those coefficients are presented in
appendix \ref{appad}, and one can notice that all resultant
$C_i^c$ and $\tilde C_i^c$ are free of infrared divergence.
The first terms of
$C_1^c(\mu_W)$ and $C_2^c(\mu_W)$ are the SM contribution whereas
the other terms are due to the supersymmetry contributions.
Other non-zero Wilson
coefficients $C_i^c(\mu_W),\tilde C_j^c(\mu_W)\;(i=3,4,...,10,
j=1,2,...,10)$ all originate from contributions of scalar quarks
and gluino.
Provided $m_{\tilde g}=m_{_{\tilde Q}}=m_{_{SUSY}}$, we have
\begin{eqnarray}
&&C_2^c(\mu_W)\sim C_{2_{\rm SM}}^c(\mu_W)=1-{\alpha_s\over
4\pi}(\ln x_{\mu}+{11\over 6}),\nonumber \\
&&C_1^c(\mu_W)\sim C_{1_{\rm SM}}^c(\mu_W)=
{3\alpha_s\over 4\pi}(\ln x_{\mu}+{11\over
6}),\;\;\;\;{\rm as}\;m_{_{SUSY}}\gg\mu_W,
\end{eqnarray}
which recovers the SM result.
When $m_{\tilde g}\rightarrow\infty$, but
$m_{\tilde U}=m_{\tilde{U}^I_i}\; (I=1,\;2,\;3;\; i=1,\;2)$
and $m_{\tilde D}=m_{\tilde{D}^I_i}\; (I=1,\;2,\;3;\; i=1,\;2)
$ remain finite, we also have $C_{1,2}^c(\mu_W)\sim
C_{1,2_{\rm SM}}^c(\mu_W)$.
From the equations given in the appendix, one observes that in
$C_2^c(\mu_W)$ if all
the mass parameters of the supersymmetric particles tend to infinity or they
are the same, i.e. the supersymmetric particles are degenerate in mass
(it is exactly the mSUGRA case) all
the $x$-value related terms cancel each other.
Here we take into account the fact
${\cal  Z}_{\tilde{U}^{1,2}}\sim
{\cal  Z}_{\tilde{D}^{1,2,3}}\sim {\bf I}$ in the mSUGRA
scenario. In other coefficients $C_1^c, \;C_3^c$ etc.
the logarithm-related terms are
automatically suppressed to zero when the supersymmetric
particles are very heavy (see appendix A for details).
Those results indicate
the decoupling of the supersymmetric sector as the supersymmetric partners
turn to be very heavy.

In order to give a complete $|\Delta B|=1$ non-leptonic effective
Lagrangian with five quarks, we should include the
contributions of penguin diagrams. In our present work, we only consider
the gluon-penguin.

\subsection{The Wilson coefficients of penguin-induced operators}

\begin{figure}
\setlength{\unitlength}{1mm}
\begin{center}
\begin{picture}(230,200)(55,90)
\put(50,-50){\includegraphics{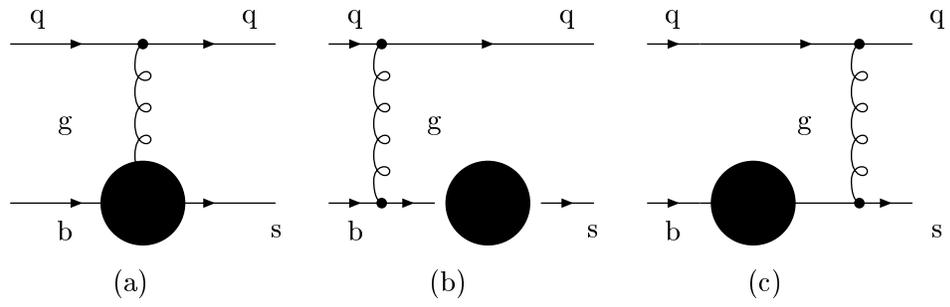}}
\end{picture}
\caption[]{The one-loop diagrams for calculating the
penguin-induced four-quark operators}
\label{fig3}
\end{center}
\end{figure}

In this section, we derive the Wilson coefficients for the
penguin-induced dimension six operators.
The basis for the penguin-induced operators is given in Eq.\ref{penguin}.
The one-loop Feynman diagrams for the penguin-induced operators
are drawn in Fig.\ref{fig3}. The obtained coefficients
$C_i^p(\mu_{\rm W})$ read
\begin{eqnarray}
&&C_1^p(\mu_{\rm W})=\frac{\alpha_s}{4\pi}\Big[-\frac{1}{9}\ln x_{\mu}
-\frac{1}{6}
E(x_t,x_{_H},x_{_{\tilde{U}_i^3}},x_{\chi_j})
+\frac{1}{9}\Big]\;,\nonumber \\
&&C_2^p(\mu_{\rm W})=\frac{\alpha_s}{4\pi}\Big[\frac{1}{3}\ln x_{\mu}+
\frac{1}{2}
E(x_t,x_{_H},x_{_{\tilde{U}_i^3}},x_{\chi_j})
-\frac{1}{3}\Big]\;,\nonumber \\
&&C_3^p(\mu_{\rm W})=\frac{\alpha_s}{4\pi}\Big[-\frac{1}{9}\ln x_{\mu}
-\frac{1}{6}
E(x_t,x_{_H},x_{_{\tilde{U}_i^3}},x_{\chi_j})
+\frac{1}{9}\Big]\;,\nonumber \\
&&C_4^p(\mu_{\rm W})=\frac{\alpha_s}{4\pi}\Big[\frac{1}{3}\ln x_{\mu}+
\frac{1}{2}
E(x_t,x_{_H},x_{_{\tilde{U}_i^3}},x_{\chi_j})
-\frac{1}{3}\Big]\;,\nonumber \\
\label{wilpeng}
\end{eqnarray}
where
\begin{eqnarray}
&&E(x_t,x_{_H},x_{_{\tilde{U}_i^3}},x_{\chi_j})=\Big[\frac{18x_t-11x_t^2
-x_t^3}{12(1-x_t)^3}+\frac{(-4+16x_t-9x_t^2)\ln x_t}{6(1-x_t)^4}\Big]
\nonumber \\
&&\hspace{3.5cm}+\frac{1}{\tan^2\beta}\Big[\frac{(2x_tx_{_H}^3-3x_t^2x_{_H}^2)
(\ln x_t-\ln x_{_H})}{6(x_{_H}-x_t)^4}+\frac{16x_tx_{_H}^2-29x_t^2x_{_H}+7x_t^3}
{36(x_{_H}-x_t)^3}\Big]
\nonumber \\
&&\hspace{3.5cm}+\sum\limits_{ij}
|{\cal A}^{ij}|^2\Big[\frac{x_{\chi_j}^3(\ln x_{_{\tilde{U}_i^3}}
-\ln x_{\chi_j})}{6(x_{\chi_j}-x_{_{\tilde{U}_i^3}})^4}+\frac{11x_{\chi_j}^2
-7x_{\chi_j}x_{_{\tilde{U}_i^3}}+2x_{_{\tilde{U}_i^3}}^2}{18(x_{\chi_j}
-x_{_{\tilde{U}_i^3}})^3}\Big]\;.
\label{efun}
\end{eqnarray}
The Wilson coefficient of $Q_5^p$ is
\begin{eqnarray}
&&C_5^p(\mu_{\rm W})=x_t\Big[\frac{3x_t\ln x_t}{4(1-x_t)
^4}+\frac{2+5x_t-x_t^2}{8(1-x_t)^3}\Big]+\bigg\{
\Big[\frac{x_tx_{_H}^2(\ln x_t-\ln x_{_H})}
{2(x_{_H}-x_t)^3}+\frac{3x_{_H}x_t-x_t^2}{4(x_{_H}-x_t)^2}\Big]
\nonumber \\
&&\hspace{2.cm}
+\frac{1}{\tan^2\beta}\Big[-\frac{x_t^2x_{_H}^2(\ln x_t-\ln x_{_H})}
{4(x_{_H}-x_t)^4}
-\frac{2x_tx_{_H}^2+5x_t^2x_{_H}-x_t^3}{24(x_{_H}-x_t)^3}\Big]
\bigg\}
\nonumber \\
&&\hspace{2.cm}+\bigg\{
\sum\limits_{ij}|{\cal A}^{ij}|^2\Big[\frac{x_{\chi_j}^2
x_{_{\tilde{U}_i^3}}(\ln x_{_{\tilde{U}_i^3}}
-\ln x_{\chi_j})}{2(x_{\chi_j}-x_{_{\tilde{U}_i^3}})^4}
+\frac{2x_{\chi_j}^2+5x_{\chi_j}x_{_{\tilde{U}_i^3}}
-x_{_{\tilde{U}_i^3}}^2}{12(x_{\chi_j}-x_{_{\tilde{U}_i^3}})^3}\Big]
\nonumber \\
&&\hspace{2.cm}
-\sum\limits_{ij}m_{\chi_j}{\cal B}^{ij}
\Big[\frac{x_{\chi_j}x_{_{\tilde{U}_i^3}}
(\ln x_{_{\tilde{U}_i^3}}-\ln x_{\chi_j})}
{(x_{\chi_j}-x_{_{\tilde{U}_i^3}})^3}
+\frac{x_{\chi_j}+x_{_{\tilde{U}_i^3}}}
{2(x_{\chi_j}-x_{_{\tilde{U}_i^3}})^2}\Big]\bigg\}.
\label{coep5}
\end{eqnarray}
The first terms of Eq.\ref{efun} and Eq.\ref{coep5} are the SM
contributions\cite{Inami,Grigjanis, Grinstein,Cella,Misiak},
the second and the
third terms are the contributions from charged Higgs and
supersymmetric particles respectively\cite{bertolini}.
In the above expression,
we only keep the contribution of
the up-type scalar quarks of the third generation, and that
from other squarks are ignored for their large masses.
The matrix ${\cal A,\;B}$ are written as
\begin{eqnarray}
&&{\cal A}^{ij}=-{\cal  Z}_{\tilde{U}^3}^{1i}{\cal  Z}_{1j}^{+*}+\frac{m_t}
{\sqrt{2}m_{\rm W}\sin\beta}{\cal  Z}_{\tilde{U}^3}^{2i}{\cal  Z}_{2j}^{+*}
\nonumber \\
&&{\cal B}^{ij}=\frac{{\cal  Z}_{\tilde{U}^3}^{1i}{\cal  Z}_{2j}^-}
{\sqrt{2}m_{\rm W}\cos\beta}
\Big(-{\cal  Z}_{\tilde{U}^3}^{1i}{\cal  Z}_{1j}^++\frac{m_t}
{\sqrt{2}m_{\rm W}\sin\beta}
{\cal  Z}_{\tilde{U}^3}^{2i}{\cal  Z}_{2j}^+\Big)
\label{Apara}
\end{eqnarray}
with ${\cal  Z}_{\tilde{U}^3},\;{\cal  Z}^{+}$ are the mixing matrices
of scalar top quarks and
charginos respectively. Because we are working in the framework of
supersymmetric extension of SM, the effective vertex $b\rightarrow
sg$ must take in the contribution of supersymmetric particles, thus
the penguin-induced Lagrangian is somewhat
different from that within SM.
It is noticed that the $m_H$ related terms i.e. the last two terms
in Eq.\ref{efun} and Eq.\ref{coep5} tend to zero when
$m_H\rightarrow \infty$ and as well as $m_{_{SUSY}}\rightarrow \infty$.

When we only keep the Yukawa coupling of the top-quark to the
Higgs boson and assuming that the masses of the scalar quarks
except top scalar quarks are highly degenerate, and the weak
eigenstates are the eigenstates of the masses, i.e.
${\cal  Z}_{\tilde U}={\cal  Z}_{\tilde D}=1$,
we find that only the coefficients of $Q_1^c,\; Q_2^c$ and
the five penguin-induced operators are not zero, whereas the Wilson
coefficients of other operators vanish. This is exactly the
operator basis existing in the standard model. In the $\overline{
\rm MS}$-scheme, the Wilson coefficients are simplified as
\begin{eqnarray}
&&C_1^c(\mu_{\rm W})=3\frac{\alpha_s}{4\pi}\Big(
\ln x_{\mu}+\frac{11}{6}\Big)-2\frac{\alpha_s}{4\pi}
\sum\limits_{k}{\cal  Z}_{1k}^{-*}{\cal  Z}_{1k}^{-}
\Big[\frac{x_{_{\tilde g}}^2\ln x_{_{\tilde g}}}
{(x_{_{\chi_k^+}}-x_{_{\tilde g}})(x_{_{\tilde Q}}-
x_{_{\tilde g}})^2}+\frac{x_{_{\chi_k^+}}^2\ln x_{_{\chi_k^+}}}
{(x_{_{\tilde g}}-x_{_{\chi_k^+}})(x_{_{\tilde Q}}
-x_{_{\chi_k^+}})^2}\nonumber \\
&&\hspace{2.cm}+\frac{x_{_{\tilde Q}}^2\ln x_{_{\tilde Q}}}
{(x_{_{\tilde g}}-x_{_{\tilde Q}})^2(x_{_{\chi_k^+}}-
x_{_{\tilde Q}})}+\frac{x_{_{\tilde Q}}^2\ln x_{_{\tilde Q}}}
{(x_{_{\tilde g}}-x_{_{\tilde Q}})(x_{_{\chi_k^+}}-
x_{_{\tilde Q}})^2}+\frac{x_{_{\tilde Q}}(2\ln x_{_{\tilde Q}}+1)}
{(x_{_{\tilde g}}-x_{_{\tilde Q}})(x_{_{\chi_k^+}}-
x_{_{\tilde Q}})}\Big]\;,
\nonumber \\
&&C_2^c(\mu_{\rm W})=\Big(1-\frac{\alpha_s}{4\pi}
(\ln x_{\mu}+\frac{11}{6})\Big)-\frac{\alpha_s}{2\pi}
\bigg(\frac{x_{_{\tilde Q}}-x_{_{\tilde Q}}\ln x_{_{\tilde Q}}}
{x_{_{\tilde g}}-x_{_{\tilde Q}}}+\frac{(x_{_{\tilde g}}
x_{_{\tilde Q}}-x_{_{\tilde Q}}^2)\ln x_{_{\tilde Q}}
+x_{_{\tilde Q}}^2\ln x_{_{\tilde g}}}{(x_{_{\tilde g}}
-x_{_{\tilde Q}})^2}\bigg)\;,
\label{simp}
\end{eqnarray}
the Wilson coefficients of the penguin-induced operators remain unchanged
and the other Wilson coefficients vanish.

\subsection{The evolution of the Wilson coefficients\label{sec4}}

Many flavor-changing processes occur at the hadronic scale with
$\mu_{hadron}\ll
\mu_{\rm W}$. Generally, first a few terms in the
perturbative expansion of the amplitudes are sufficient
when the renormalization scale $\mu$ is close to $\mu_{hadron}$ rather
than to $\mu_{\rm W}$\cite{Chetyrkin}. The Wilson coefficients $C_i^c,\;
\tilde{C}_i^c,\; C_i^p$ at $\mu_{hadron}$ are obtained from $C_i^c(\mu_{\rm W})
,\;\tilde{C}_i^c(\mu_{\rm W}),\;C_i^p(\mu_{\rm W})$ with help of the
Renormalization Group Equations (RGEs) evolution. If we define a $1\times 24$
matrix as
\begin{equation}
\vec{\bf C}=\Big(C_1^c,C_2^c,\cdots,C_{10}^c,\tilde{C}_1^c,\tilde{C}_2^c,
\cdots,\tilde{C}_{10}^c,C_1^p,C_2^p,C_3^p,C_4^p\Big)\;,
\label{wmatr}
\end{equation}
and the RGEs for the Wilson coefficients are
\begin{equation}
\mu\frac{d}{d\mu}\vec{\bf C}(\mu)=\vec{\bf C}(\mu)\hat{\gamma}(\mu)\;.
\label{wrge}
\end{equation}
Here, $\hat{\gamma}$ is the anomalous dimension matrix which has
the following form in the perturbative expansion
\begin{equation}
\hat{\gamma}(\mu)=\frac{\alpha_s(\mu)}{4\pi}\hat{\gamma}^{(0)}
+\Big(\frac{\alpha_s(\mu)}{4\pi}\Big)^2\hat{\gamma}^{(1)}.
\label{anom}
\end{equation}
Adopting the Naive Dimensional Regularization$-\overline {\rm MS}$
(NDR-$\overline{\rm MS}$) scheme, Buras {\it et.al} have
given the anomalous dimension matrix up to two-loop order in the
basis of Eq.\ref{current} and Eq.\ref{penguin}\cite{Buras1,Buras6}.

With the RGEs (\ref{wrge}) and the Wilson
coefficients at the weak scale
as the initial condition, we can derive the effective Lagrangian
of five quarks at the hadronic scale.

\section{Numerical results}

Indeed,  there are too many
free parameters in the minimal supersymmetric extension of SM (MSSM). In
order to reduce the number of free parameters, we assume that the MSSM
is a low-energy effective theory of a more fundamental theory
which exists at a higher scale, such as the grand unification scale
or the Planck scale. A realization of this idea is the
minimal supergravity (mSUGRA), which is fully specified by five
parameters\cite{Fengj}
$$m_0,\;m_\frac{1}{2},\;A_0,\;\tan\beta,\;sgn(\mu).$$
Here $m_0,\;m_{\frac{1}{2}}$ and $A_0$ are the universal
scalar quark mass, gaugino mass and trilinear scalar coupling. They are
assumed to arise through supersymmetry breaking in a hidden-sector
at the GUT scale $\mu_{GUT}\simeq 2\times 10^{16}$GeV. In our
numerical calculation, to maintain consistency of the theory and
the up-to-date experimental observation, when we obtain the
numerical value of the Higgs mass in the mSUGRA model with
the five parameters, we include all one-loop effects in the
Higgs potential\cite{Pierce}. Moreover  we also employ the two-loop
RGEs\cite{2loop} with one-loop threshold
corrections\cite{Pierce,Bagger1} as the energy scale runs
down from the mSUGRA scale to the lower weak scale.
In the framework of minimal supergravity, the unification assumptions
are expressed as
\begin{eqnarray}
&&A_l^I=A_d^I=A_u^I=A_0\;,\\ \nonumber
&&B=A_0-1\;, \\ \nonumber
&&m_{H^1}^2=m_{H^2}^2=m_{L^I}^2=m_{R^I}^2=m_{Q^I}^2=m_{U^I}^2=
m_{D^I}^2=m_0^2\;,\\ \nonumber
&&m_1=m_2=m_3=m_{\frac{1}{2}}\;.
\label{unifi}
\end{eqnarray}
For the SM parameters, we have $m_b=5{\rm GeV},\; m_t=174{\rm GeV}
,\; m_{\rm W}=80.23{\rm GeV},\;\alpha_e(m_{\rm W})=\frac{1}{128},\;
\alpha_s(m_{\rm W})=0.12$ at the weak scale. Taking above values, we find
the SM prediction for Wilson Coefficients as $C_{1_{SM}}^{c}(m_b)=-0.295
,\;C_{2_{SM}}^{c}(m_b)=1.110,\;C_{1_{SM}}^{p}(m_b)=0.014$.
In our numerical calculations of the supersymmetry corrections to those Wilson
coefficients $C_i^c(m_b)\;(i=1,...,10),\;\tilde C_i^c(m_b)\;(i=1,...,10)$,
and $C_i^p(m_b)$,
we always set $A_0=0$ and $sgn(\mu)=+$. Even though other Wilson
coefficients also get nonzero contributions from the supersymmetric
sector, our discussions mainly focus at the dependence of
$C_1^c(m_b),\;C_2^c(m_b),\;C_{1}^p(m_b)$ on the
supersymmetric parameters  because
they play more significant roles in low energy
phenomenology. In Fig.\ref{fig4} (a), (b) and (c), we plot the
ratios between supersymmetry corrections to
$C_1^c(m_b),\;C_2^c(m_b),\;C_1^p(m_b)$
and their SM prediction
values versus the parameter $m_{\frac{1}{2}}$ with $m_0=200$GeV and
$\tan\beta=2$ or $20$.
The dependence of those ratios on $m_0$
($m_{1\over 2}=300$GeV and $\tan\beta=2$ or $20$) is
plotted in Fig.\ref{fig5}.

\section{Discussions and Conclusion}

The $|\Delta B|=1$ non-leptonic effective Lagrangian has been
considered in the minimal flavor violating supersymmetry
scenario. The
supersymmetry contributions affect the effective Lagrangian
via two aspects:
\begin{itemize}
\item new current-current operators emerge
beside those 'old' operators in the SM case;
\item for the 'old' operators, the Wilson coefficients at the weak
scale are modified by the supersymmetric contributions.
\end{itemize}

Now let us briefly discuss our observation of the numerical
results.

\begin{figure}
\setlength{\unitlength}{1mm}
\begin{center}
\begin{picture}(230,200)(55,90)
\put(30,70){\includegraphics{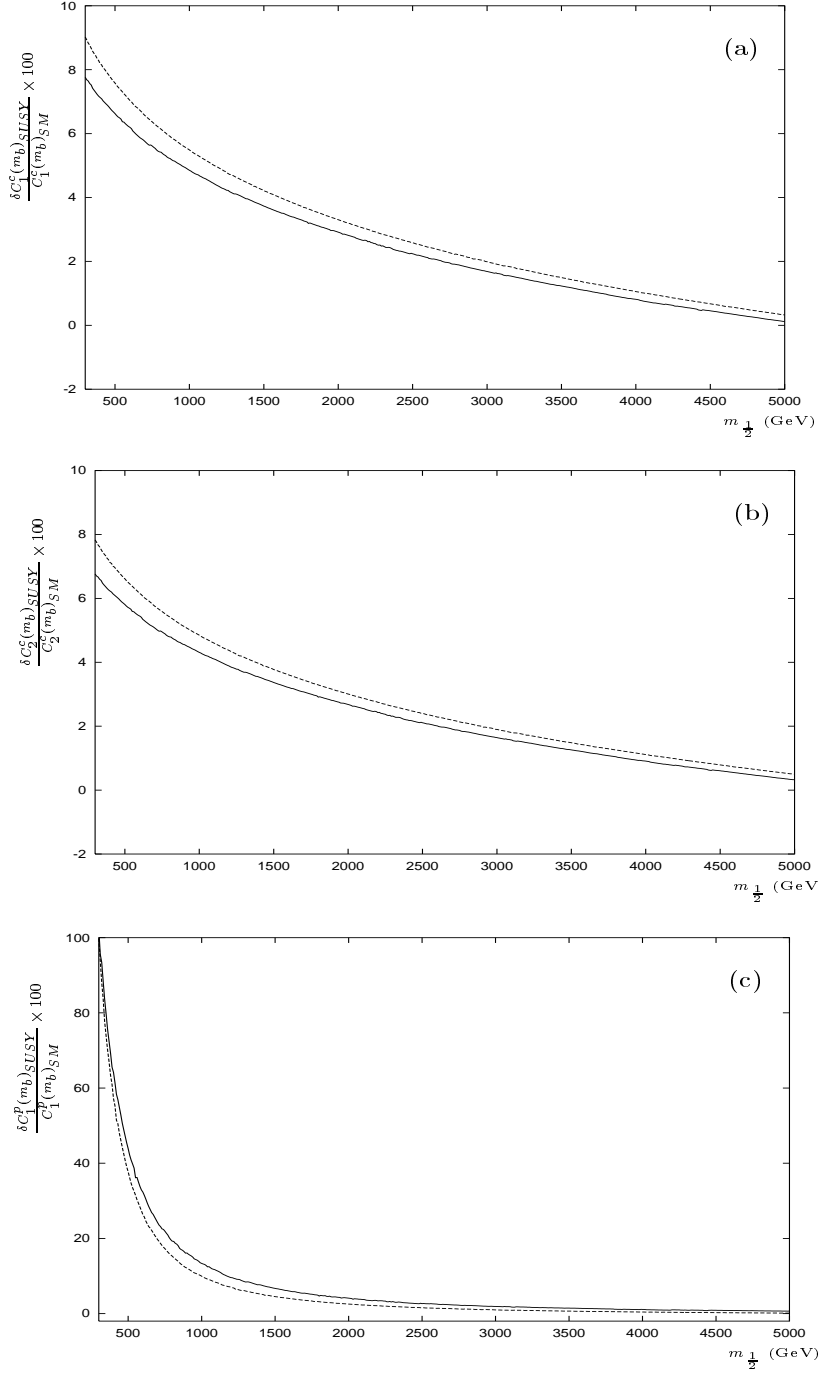}}
\end{picture}
\caption[]{The relative supersymmetry corrections
(the supersymmetric corrections/ the SM predictions) to the
Wilson coefficients at the $m_b$ scale versus $m_\frac{1}{2}$ with
$\tan\beta=2$ (Solid-lines), $\tan\beta=20$
(Dash-Lines). The other parameters are set as $m_0=200{\rm GeV}
,\; A_0=0,\; sgn(\mu)=+$.}
\label{fig4}
\end{center}
\end{figure}

\begin{figure}
\setlength{\unitlength}{1mm}
\begin{center}
\begin{picture}(230,200)(55,90)
\put(30,70){\includegraphics{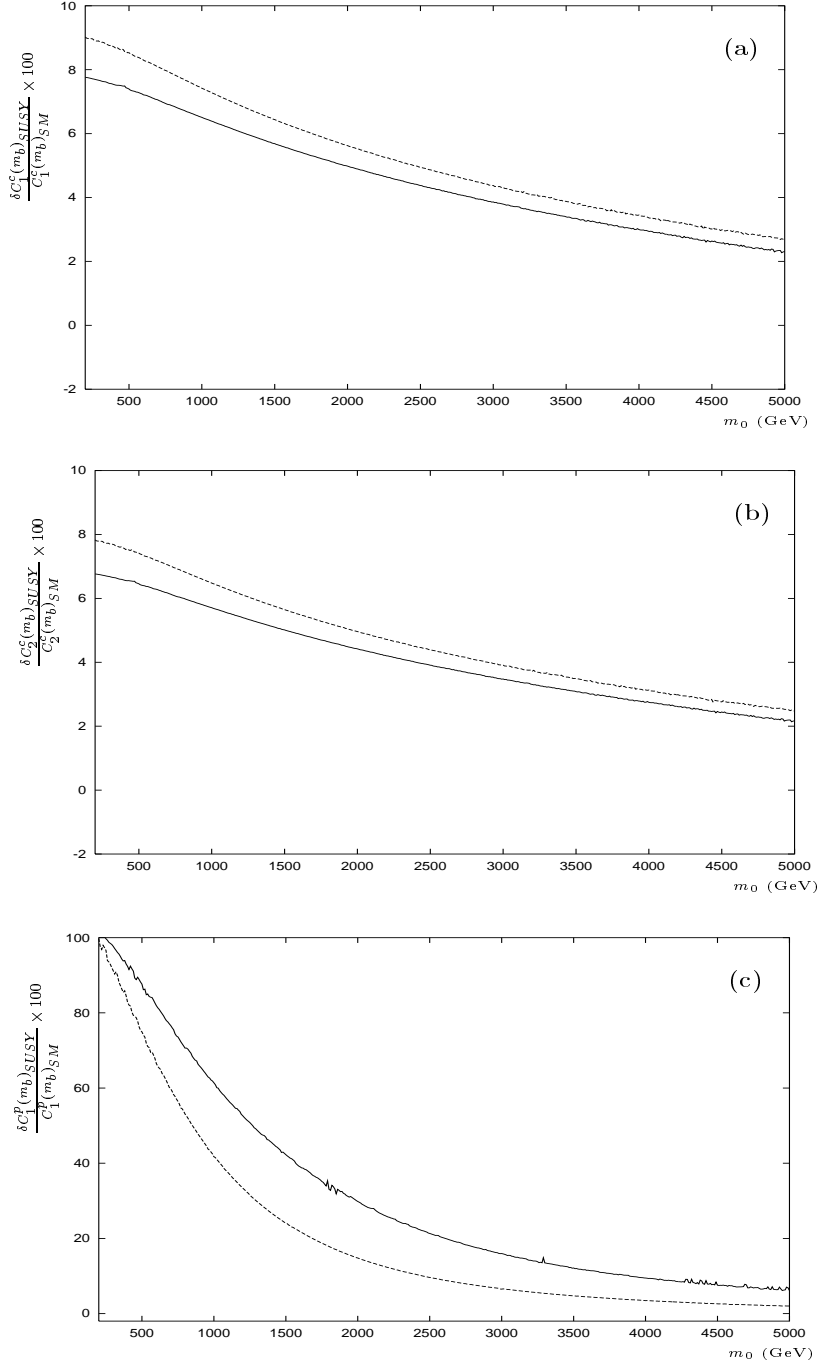}}
\end{picture}
\caption[]{The relative supersymmetry corrections to the
Wilson Coefficients at the $m_b$ scale versus
$m_0$ with $\tan\beta=2$ (Solid-lines), $\tan\beta=20$
(Dash-Lines). The other parameters are set as $m_\frac{1}{2}=300{\rm GeV}
,\;A_0=0,\;sgn(\mu)=+$.}
\label{fig5}
\end{center}
\end{figure}

In Fig.\ref{fig4} (a) and (b), the two lines (solid and dash) corresponding to
$\tan\beta=2$ and $\tan\beta=20$ differ from each other more obviously.
The supersymmetry corrections to the Wilson coefficients $C_{1,2}^c(m_b)$
are relatively large when the supersymmetry particles have masses
of the same order of electroweak energy scale, for example,
$m_{1\over 2}=300{\rm GeV},\;m_0=200{\rm GeV}$, the
supersymmetry corrections to $C_{1,2}^c(m_b)$ can reach about 8\%.
When the  masses of the supersymmetric particles become very large,
the supersymmetry corrections turn to zero due to the
decoupling theorem. In Fig.\ref{fig4}(c), the two lines
corresponding to $\tan\beta=2$ and $\tan\beta=20$ almost overlap
on each other. It is noted that at the left part of Fig.5 (c)
as $m_{1\over 2}< 2$ TeV, there is a sharp peak at the dependence
of $C_1^p(m_b)$ on $m_{1\over 2}$. It drops very fast as
$m_{1\over 2}$ is away from
this region, the resonance is due to an almost degeneracy of
the mass of chargino and the mass of stop and it leads to an
obvious deviation of the $C_{1}^p(m_b)$ value from the prediction
of SM (0.014). As $m_{1\over 2}$ further increases, $C_1^p$
tends to the predictive value of SM.  This mass resonance only
occurs for the coefficients of the penguin-induced
operators, but not for the current-current-quark
operators, so that the dependence of $C_1^c(m_b), C_2^c(m_b)$ on
$m_{1\over 2}$ does not change drastically.
All these are consistent with
the common sense which is familiar to us even before the
calculations are done.

In Fig.\ref{fig5}, we plot the dependence of $C_1^c(m_b), C_2^c(m_b)$ and
$C_1^p$ on $m_0$, with $A_0=0,\;sgn(\mu)=+1,\; m_{1\over 2}=200$ GeV.
Similar to the case of Fig.\ref{fig4}, the dependence of supersymmetry
corrections to $C_{1,2}^c(m_b)$ on $m_0$ is remarkable and the
values obviously deviate from the SM prediction when $m_0$ take
smaller values. When $m_0\sim 200{\rm GeV}$, $C_{1,2}^c(m_b)$ deviate from
the SM prediction by 8\%, and when $m_0>5 $ TeV, the deviation tends
to zero. The dependence of $C_1^p(m_b)$ on $m_0$ is in analog to its
dependence on $m_{1\over 2}$. When $m_0< 2$ TeV, there is an obvious peak
which damps steeply,
the reason is still due to the mass degeneracy of heavy chargino and stop.
When $m_0$ turns very large, $C_1^p(m_b)$ approaches the SM
prediction. By contrast, the dependence of $C_1^c(m_b),
C_2^c(m_b)$ on $m_0$ is more smooth, because there is no mass resonance
effect in this situation.

In above discussions, we only list the Wilson coefficients of
a few typical operators, the Wilson coefficients of other
operators are in analogy. For simplicity, in all the numerical
calculations, we always adopt $A_0=0,\;sgn(\mu)=+1$, but this
convention is not necessary. If we dismiss this assumption and let
$A_0\neq 0,\; sgn(\mu)=\pm 1$, the parameter space is enlarged,
but the qualitative conclusion remains the same.

Here we briefly discuss the decoupling of the supersymmetry particles as the
concerned energy scale turns to infinity.
%\footnote{Here we would
%like to thank the referee who read our first version and made
%comments on this issue, his suggestions led to some radical
%modifications to the work.}
When we match the effective theory to the full theory which is
the standard model at the weak scale, the quark and gluon fields
exist in both theories, thus the renormalized quark fields are
the same in both theories. However, as the full theory is the
supersymmetric extension of SM, scalar quarks and gluinos exist only in
the full theory, but do not in the effective one. Thus when
we match the two theories at a certain scale, disappearance
of such supersymmetric partners in the effective theory would
result in a difference of the normalization of the quark fields
in the full and effective theories. Actually, this is a finite
renormalization contribution of  the self-energy.
Ignoring this effect would lead to
a situation that heavy supersymmetry particles do not decouple. This
normalization difference would also affect evaluation of the
vertex-induced contribution. When we match the full and effective
theories we need to take
this normalization difference into account,
then the large logarithms emerging from the
self-energy exactly cancel out that from the vertices, thus
the decoupling is obvious.
On other side, only in the $\overline{\rm MS}$ scheme, there exists
the difference of the normalization of the renormalized quark fields,
but it does not exist in the on-mass shell renormalization scheme.
The reason is that requiring the external
quarks to be on their physical mass shell and the residue for the
self-energy to be one can serve as an additional condition by
which the normalization of the quark fields are the same in the
full and effective theories. A direct consequence of the correct
renormalization scheme is the decoupling of the supersymmetry sector as the
supersymmetry particles are too heavy.   If we include the normalization
difference as we take the $\overline{\rm MS}$ scheme, the result
coincides with that in the on-shell renormalization.

Because we have carefully considered the normalization difference,
in our final expressions,
one can immediately observe that the supersymmetry particles would
decouple if their mass scale turns to infinity. That is
consistent with the common sense.

From our discussion and numerical results, one can note that in
general, the correction of supersymmetry to the Wilson coefficients can
be as large as 8\% when the supersymmetry particles are not very
heavy. It is also noted that  the $|\Delta B|=1$
effective Lagrangian at the weak scale induces
an electromagnetic dipole and a
chromo-dipole. Through the evolution  down to the
hadronic scale, the supersymmetric contribution would result in substantial
values for the dipoles at low energies which
may not be negligible for phenomenology when
$\tilde g$ and $\tilde Q$ are relative light.

With collection of large amount of data at the B-factory
and elsewhere main laboratories in the world, the
measurements on the rare processes $b\rightarrow s\gamma,\;
b\rightarrow s\bar uu$ etc. would set more rigorous constraint on
the parameter space of the supersymmetric model or we can, as expected, find
some evidence of existence of supersymmetric particles.

\vspace{0.5cm}
\noindent {\Large{\bf Acknowledgments}}

This work is partially supported by the National Natural Science
Foundation of China.

\vspace{0.5cm}
\appendix

\section{The Wilson coefficients of current-current operators at the weak
scale \label{appad}}

After matching at the weak scale, the Wilson coefficients
for the current-current operators are written as
\begin{eqnarray}
&&C_1^c(\mu_{\rm W})=3\frac{\alpha_s}{4\pi}\Big(
\ln x_{\mu}+\frac{11}{6}\Big)-\frac{\alpha_s}{4\pi}
\Big[\sum\limits_{\alpha=\tilde{g},\chi_k^+,\tilde{D}_i^3,\tilde{D}_j^2}
\frac{x_{\alpha}^2\ln x_{\alpha}}{\prod\limits_{\beta\neq \alpha}
(x_{\beta}-x_{\alpha})}{\cal  Z}_{\tilde{D}^2}^{1j*}
{\cal  Z}_{\tilde{D}^3}^{1i}A_{bc}^{-ik*}A_{sc}^{-jk}
\nonumber \\
&&\hspace{2.cm}+\sum\limits_{\alpha=\tilde{g},\chi_k^+,\tilde{U}_i^2,
\tilde{U}_j^2}
\frac{x_{\alpha}^2\ln x_{\alpha}}{\prod\limits_{\beta\neq \alpha}
(x_{\beta}-x_{\alpha})}
{\cal  Z}_{\tilde{U}^2}^{1j*}{\cal  Z}_{\tilde{U}^2}^{1i}
B_{bc}^{-ik}B_{sc}^{-jk*}\Big]\;,
\nonumber \\
&&C_2^c(\mu_{\rm W})=\Big(1-\frac{\alpha_s}{4\pi}
(\ln x_{\mu}+\frac{11}{6})\Big)
-\frac{\alpha_s}{8\pi}\bigg\{2\sum\limits_{j}
Z_{\tilde{U}^2}^{1j}Z_{\tilde{U}^2}^{1j*}\Big(
\frac{x_{_{\tilde{U}_j^2}}}{x_{_{\tilde{g}}}
-x_{_{\tilde{U}_j^2}}}+\frac{x_{_{\tilde{U}_j^2}}\ln
x_{_{\tilde{U}_j^2}}}{x_{_{\tilde{g}}}-x_{_{\tilde{U}_j^2}}}\nonumber \\
&&\hspace{2cm}
-\frac{(2x_{_{\tilde{g}}}x_{_{\tilde{U}_j^2}}-x_{_{\tilde{U}_j^2}}^2
)\ln x_{_{\tilde{g}}}-x_{_{\tilde{g}}}
x_{_{\tilde{U}_j^2}}\ln x_{_{\tilde{U}_j^2}}}
{(x_{_{\tilde{g}}}-x_{_{\tilde{U}_j^2}})^2}\Big)+\sum\limits_{i}
Z_{\tilde{D}^2}^{1i}Z_{\tilde{D}^2}^{1i*}\Big(
\frac{x_{_{\tilde{D}_i^2}}}{x_{_{\tilde{g}}}
-x_{_{\tilde{D}_i^2}}}+\frac{x_{_{\tilde{D}_i^2}}\ln
x_{_{\tilde{D}_i^2}}}{x_{_{\tilde{g}}}-x_{_{\tilde{D}_i^2}}}\nonumber \\
&&\hspace{2.cm}
-\frac{(2x_{_{\tilde{g}}}x_{_{\tilde{D}_i^2}}-x_{_{\tilde{D}_i^2}}^2
)\ln x_{_{\tilde{g}}}-x_{_{\tilde{g}}}
x_{_{\tilde{D}_i^2}}\ln x_{_{\tilde{D}_i^2}}}
{(x_{_{\tilde{g}}}-x_{_{\tilde{D}_i^2}})^2}\Big)
+2\sum\limits_{ij}Z_{\tilde{D}^2}^{1i}
Z_{\tilde{D}^2}^{1i*}Z_{\tilde{U}^2}^{1j}Z_{\tilde{U}^2}^{1j*}
\Big\{\nonumber \\
&&\hspace{2.cm}\frac{\Big(x_{_{\tilde{g}}}(x_{_{\tilde{D}_i^2}}
+x_{_{\tilde{U}_j^2}})-x_{_{\tilde{D}_i^2}}
x_{_{\tilde{U}_j^2}}\Big)\ln x_{_{\tilde{g}}}}{(x_{_{\tilde{D}_i^2}}
-x_{_{\tilde{g}}})(x_{_{\tilde{U}_j^2}}-x_{_{\tilde{g}}})}
+\frac{x_{_{\tilde{D}_i^2}}^2\ln x_{_{\tilde{D}_i^2}}}
{(x_{_{\tilde{g}}}-x_{_{\tilde{D}_i^2}})(x_{_{\tilde{U}_j^2}}-
x_{_{\tilde{D}_i^2}})}\nonumber \\
&&\hspace{2.cm}+\frac{x_{_{\tilde{U}_j^2}}^2\ln x_{_{\tilde{U}_j^2}}}
{(x_{_{\tilde{g}}}-x_{_{\tilde{U}_j^2}})(x_{_{\tilde{D}_i^2}}
-x_{_{\tilde{U}_j^2}})}\Big\}+\sum\limits_{i}
Z_{\tilde{D}^3}^{1i}Z_{\tilde{D}^3}^{1i*}\Big(
\frac{x_{_{\tilde{D}_i^3}}}{x_{_{\tilde{g}}}
-x_{_{\tilde{D}_i^3}}}+\frac{x_{_{\tilde{D}_i^3}}\ln
x_{_{\tilde{D}_i^3}}}{x_{_{\tilde{g}}}-x_{_{\tilde{D}_i^3}}}\nonumber \\
&&\hspace{2.cm}
-\frac{(2x_{_{\tilde{g}}}x_{_{\tilde{D}_i^3}}-x_{_{\tilde{D}_i^3}}^2
)\ln x_{_{\tilde{g}}}-x_{_{\tilde{g}}}
x_{_{\tilde{D}_i^3}}\ln x_{_{\tilde{D}_i^3}}}
{(x_{_{\tilde{g}}}-x_{_{\tilde{D}_i^3}})^2}\Big)
+2\sum\limits_{ij}Z_{\tilde{D}^3}^{1i}
Z_{\tilde{D}^3}^{1i*}Z_{\tilde{U}^2}^{1j}Z_{\tilde{U}^2}^{1j*}
\Big\{\nonumber \\
&&\hspace{2.cm}\frac{\Big(x_{_{\tilde{g}}}(x_{_{\tilde{D}_i^3}}
+x_{_{\tilde{U}_j^2}})-x_{_{\tilde{D}_i^3}}
x_{_{\tilde{U}_j^2}}\Big)\ln x_{_{\tilde{g}}}}{(x_{_{\tilde{D}_i^3}}
-x_{_{\tilde{g}}})(x_{_{\tilde{U}_j^2}}-x_{_{\tilde{g}}})}
+\frac{x_{_{\tilde{D}_i^3}}^2\ln x_{_{\tilde{D}_i^3}}}
{(x_{_{\tilde{g}}}-x_{_{\tilde{D}_i^3}})(x_{_{\tilde{U}_j^2}}-
x_{_{\tilde{D}_i^3}})}\nonumber \\
&&\hspace{2.cm}+\frac{x_{_{\tilde{U}_j^2}}^2\ln x_{_{\tilde{U}_j^2}}}
{(x_{_{\tilde{g}}}-x_{_{\tilde{U}_j^2}})(x_{_{\tilde{D}_i^3}}
-x_{_{\tilde{U}_j^2}})}\Big\}\bigg\}\;,\nonumber \\
&&C_3^c(\mu_{\rm W})=-\frac{\alpha_s}{4\pi}\frac{2m_{\tilde{g}}m_{\chi^+_k}}
{m_{\rm W}^2}\Big(\sum\limits_{\alpha=\tilde{g},\chi_k^+,\tilde{D}_i^3,
\tilde{D}_j^2}\frac{x_{\alpha}\ln x_{\alpha}}{\prod\limits_{\beta\neq
\alpha}(x_{\beta}-x_{\alpha})}
{\cal  Z}_{\tilde{D}^2}^{2j*}{\cal  Z}_{\tilde{D}^3}^{1i}A_{bc}^{+ik*}A_{sc}^{-jk}
\nonumber \\
&&\hspace{2.cm}+\sum\limits_{\alpha=\tilde{g},\chi_k^+,\tilde{U}_i^2,
\tilde{U}_j^2}\frac{x_{\alpha}\ln x_{\alpha}}{\prod\limits_{\beta\neq \alpha}
(x_{\beta}-x_{\alpha})}
{\cal  Z}_{\tilde{U}^2}^{2i}{\cal  Z}_{\tilde{U}^2}^{1j*}B_{bc}^{-ik}B_{sc}^{+jk*}\Big)\;,
\nonumber \\
&&C_4^c(\mu_{\rm W})=0\;, \nonumber \\
&&C_5^c(\mu_{\rm W})=-\frac{1}{4}C_3^c(\mu_{\rm W})\;,\nonumber \\
&&C_6^c(\mu_{\rm W})=0\;, \nonumber \\
&&C_7^c(\mu_{\rm W})=-\frac{\alpha_s}{4\pi}\frac{2m_{\tilde{g}}m_{\chi^+_k}}
{m_{\rm W}^2}\Big(\sum\limits_{\alpha=\tilde{g},\chi_k^+,\tilde{D}_i^3,
\tilde{D}_j^2}\frac{x_{\alpha}\ln x_{\alpha}}
{\prod\limits_{\beta\neq \alpha}(x_{\beta}-x_{\alpha})}
{\cal  Z}_{\tilde{D}^2}^{1j*}{\cal  Z}_{\tilde{D}^3}^{2i}A_{bc}^{+ik*}A_{sc}^{-jk}
\nonumber \\
&&\hspace{2.cm}+\sum\limits_{\alpha=\tilde{g},\chi_k^+,\tilde{U}_i^2,
\tilde{U}_j^2}\frac{x_{\alpha}\ln x_{\alpha}}{\prod\limits_{\beta\neq \alpha}
(x_{\beta}-x_{\alpha})}{\cal  Z}_{\tilde{U}^2}^{2i}{\cal  Z}_{\tilde{U}^2}^{1j*}
B_{bc}^{+ik}B_{sc}^{-jk*}\Big)\;,
\nonumber \\
&&C_8^c(\mu_{\rm W})=-\frac{\alpha_s}{4\pi}\sum\limits_{\alpha=\tilde{g},
\tilde{U}_i^2,\tilde{D}_j^3}\frac{x_{\alpha}^2\ln x_{\alpha}}
{\prod\limits_{\beta\neq \alpha}(x_{\beta}-x_{\alpha})}
{\cal  Z}_{\tilde{U}^2}^{2i}{\cal  Z}_{\tilde{U}^2}^{1i*}
{\cal  Z}_{\tilde{D}^3}^{2j}{\cal  Z}_{\tilde{D}^3}^{1j*}\;,
\nonumber \\
&&C_9^c(\mu_{\rm W})=2\frac{\alpha_s}{4\pi}
\Big[\sum\limits_{\alpha=\tilde{g},\chi_k^+,\tilde{D}_i^3,\tilde{D}_j^2}
\frac{x_{\alpha}^2\ln x_{\alpha}}{\prod\limits_{\beta\neq \alpha}
(x_{\beta}-x_{\alpha})}{\cal  Z}_{\tilde{D}^2}^{2j*}
{\cal  Z}_{\tilde{D}^3}^{2i}A_{bc}^{-ik*}A_{sc}^{-jk}
\nonumber \\
&&\hspace{2.cm}+\sum\limits_{\alpha=\tilde{g},\chi_k^+,\tilde{U}_i^2,
\tilde{U}_j^2}
\frac{x_{\alpha}^2\ln x_{\alpha}}{\prod\limits_{\beta\neq \alpha}
(x_{\beta}-x_{\alpha})}
{\cal  Z}_{\tilde{U}^2}^{1j*}{\cal  Z}_{\tilde{U}^2}^{1i}B_{bc}^{+ik}B_{sc}^{+jk*}\Big]\;,
\nonumber \\
&&C_{10}^c(\mu_{\rm W})=0\;,\nonumber \\
&&\tilde{C}_1^c(\mu_{\rm W})=-\frac{\alpha_s}{4\pi}
\Big[\sum\limits_{\alpha=\tilde{g},\chi_k^+,\tilde{D}_i^3,\tilde{D}_j^2}
\frac{x_{\alpha}^2\ln x_{\alpha}}{\prod\limits_{\beta\neq \alpha}
(x_{\beta}-x_{\alpha})}{\cal  Z}_{\tilde{D}^2}^{2j*}
{\cal  Z}_{\tilde{D}^3}^{2i}A_{bc}^{+ik*}A_{sc}^{+jk}
\nonumber \\
&&\hspace{2.cm}+\sum\limits_{\alpha=\tilde{g},\chi_k^+,\tilde{U}_i^2,
\tilde{U}_j^2}
\frac{x_{\alpha}^2\ln x_{\alpha}}{\prod\limits_{\beta\neq \alpha}
(x_{\beta}-x_{\alpha})}
{\cal  Z}_{\tilde{U}^2}^{2j*}{\cal  Z}_{\tilde{U}^2}^{2i}B_{bc}^{+ik}B_{sc}^{+jk*}\Big]\;,
\nonumber \\
&&\tilde{C}_2^c(\mu_{\rm W})=0\;,\nonumber \\
&&\tilde{C}_3^c(\mu_{\rm W})=-\frac{\alpha_s}{4\pi}\frac{2m_{\tilde{g}}
m_{\chi^+}}
{m_{\rm W}^2}\Big(\sum\limits_{\alpha=\tilde{g},\chi_k^+,\tilde{D}_i^3,
\tilde{D}_j^2}\frac{x_{\alpha}\ln x_{\alpha}}
{\prod\limits_{\beta\neq \alpha}(x_{\beta}-x_{\alpha})}
{\cal  Z}_{\tilde{D}^2}^{1j*}{\cal  Z}_{\tilde{D}^3}^{2i}A_{bc}^{-ik*}A_{sc}^{+jk}
\nonumber \\
&&\hspace{2.cm}+\sum\limits_{\alpha=\tilde{g},\chi_k^+,\tilde{U}_i^2,
\tilde{U}_j^2}\frac{x_{\alpha}\ln x_{\alpha}}{\prod\limits_{\beta\neq \alpha}
(x_{\beta}-x_{\alpha})}
{\cal  Z}_{\tilde{U}^2}^{1i}{\cal  Z}_{\tilde{U}^2}^{2j*}B_{bc}^{+ik}B_{sc}^{-jk*}\Big)\;,
\nonumber \\
&&\tilde{C}_4^c(\mu_{\rm W})=0\;, \nonumber \\
&&\tilde{C}_5^c(\mu_{\rm W})=-\frac{1}{4}\tilde{C}_3^c(\mu_{\rm W})
\;,\nonumber \\
&&\tilde{C}_6^c(\mu_{\rm W})=0\;, \nonumber \\
&&\tilde{C}_7^c(\mu_{\rm W})=-\frac{\alpha_s}{4\pi}
\frac{2m_{\tilde{g}}m_{\chi^+}}
{m_{\rm W}^2}\Big(\sum\limits_{\alpha=\tilde{g},\chi_k^+,\tilde{D}_i^3,
\tilde{D}_j^2}\frac{x_{\alpha}\ln x_{\alpha}}
{\prod\limits_{\beta\neq \alpha}(x_{\beta}-x_{\alpha})}
{\cal  Z}_{\tilde{D}^2}^{2j*}{\cal  Z}_{\tilde{D}^3}^{1i}A_{bc}^{-ik*}A_{sc}^{+jk}
\nonumber \\
&&\hspace{2.cm}+\sum\limits_{\alpha=\tilde{g},\chi_k^+,\tilde{U}_i^2,
\tilde{U}_j^2}\frac{x_{\alpha}\ln x_{\alpha}}
{\prod\limits_{\beta\neq \alpha}(x_\beta-x_\alpha)}
{\cal  Z}_{\tilde{U}^2}^{1i}{\cal  Z}_{\tilde{U}^2}^{2j*}B_{bc}^{-ik}B_{sc}^{+jk*}\Big)\;,
\nonumber \\
&&\tilde{C}_8^c(\mu_{\rm W})=-\frac{\alpha_s}{4\pi}
\sum\limits_{\alpha=\tilde{g},
\tilde{U}_i^2,\tilde{D}_j^3}\frac{x_\alpha^2\ln x_\alpha}
{\prod\limits_{\beta\neq \alpha}
(x_\beta-x_\alpha)}{\cal  Z}_{\tilde{U}^2}^{1i}{\cal  Z}_{\tilde{U}^2}^{2i*}
{\cal  Z}_{\tilde{D}^3}^{1j}{\cal  Z}_{\tilde{D}^3}^{2j*}\;,
\nonumber \\
&&\tilde{C}_9^c(\mu_{\rm W})=2\frac{\alpha_s}{4\pi}
\Big[\sum\limits_{\alpha=\tilde{g},\chi_k^+,\tilde{D}_i^3,\tilde{D}_j^2}
\frac{x_\alpha^2\ln x_\alpha}{\prod\limits_{\beta\neq \alpha}
(x_\beta-x_\alpha)}{\cal  Z}_{\tilde{D}^2}^{1j*}
{\cal  Z}_{\tilde{D}^3}^{1i}A_{bc}^{+ik*}A_{sc}^{+jk}
\nonumber \\
&&\hspace{2.cm}+\sum\limits_{\alpha=\tilde{g},\chi_k^+,
\tilde{U}_i^2,\tilde{U}_j^2}
\frac{x_\alpha^2\ln x_\alpha}{\prod\limits_{\beta\neq \alpha}
(x_\beta-x_\alpha)}
{\cal  Z}_{\tilde{U}^2}^{2j*}{\cal  Z}_{\tilde{U}^2}^{2i}B_{bc}^{-ik}B_{sc}^{-jk*}\Big]\;,
\nonumber \\
&&\tilde{C}_{10}^c(\mu_{\rm W})=0\;.
\label{wilsoncoe}
\end{eqnarray}
The notation $A_{IJ}^{\mp ij},\;
B_{IJ}^{\mp, ij}$ ($I,\;J=1,\;2,\;3$ are the indices of generations) are defined as
\begin{eqnarray}
&&A_{IJ}^{-ij}=-\Big({\cal Z}_{\tilde{D}^I}^{1i}
{\cal Z}_{1j}^--\frac{m_{d^I}}{\sqrt{2}m_{\rm W}\cos\beta}
{\cal Z}_{\tilde{D}^I}^{2i}{\cal Z}_{2j}^-\Big)\;,\nonumber \\
&&A_{IJ}^{+ij}=\frac{m_{u^J}}
{\sqrt{2}m_{\rm W}\sin\beta}{\cal Z}_{\tilde{D}^I}^{1i}
{\cal Z}_{2j}^{+*}\;,\nonumber \\
&&B_{IJ}^{-ij}=-\Big({\cal Z}_{\tilde{U}^I}^{1i*}
{\cal Z}_{1j}^+-\frac{m_{u^J}}{\sqrt{2}m_{\rm W}\sin\beta}
{\cal Z}_{\tilde{U}^J}^{2i*}{\cal Z}_{2j}^+\Big)\;,\nonumber \\
&&B_{IJ}^{+ij}=\frac{m_{d^I}}{\sqrt{2}m_{\rm W}\cos\beta}
{\cal Z}_{\tilde{U}^J}^{1i*}{\cal Z}_{2j}^{-*}\;.
\label{ab}
\end{eqnarray}
Here, ${\cal Z}^{\pm}$ are the mixing matrices of charginos.
%%%%%%%%%%%%%%=============================>

%\end{center}
%\begin{center}
%\end{center}
%\begin{center}
%\end{center}
%\begin{center}
%\end{center}
%\begin{center}
%\end{center}
%\begin{center}

\end{document}